\documentclass[proceedings]{JHEP3}

\PrHEP{PrHEP hep2001}
\conference{International Europhysics Conference on HEP}

\usepackage{epsfig}
\usepackage{amssymb}

\newcommand{\ppbar}{$p\bar{p}$}
\newcommand{\ccbar}{$c\bar{c}$}
\newcommand{\bbbar}{$b\bar{b}$}
\newcommand{\Lepto}{\textsc{lepto}}
\newcommand{\Pythia}{\textsc{pythia}}

\title{Soft Colour Interactions and \\
Diffractive Hard Scattering at the Tevatron}

\author{\speaker{Nicusor T\^\i mneanu},                       
        Rikard Enberg and Gunnar Ingelman\\  
        High Energy Physics, Uppsala University, 
	Box 535, S-751~21 Uppsala, Sweden\\                                          
        E-mail: \email{nicusor@tsl.uu.se, enberg@tsl.uu.se, ingelman@tsl.uu.se}}                       

\abstract{We make a brief presentation of the soft colour interactions
models, the Soft Colour Interaction and the Generalised Area Law, 
and summarise the results when they are applied to \ppbar~scattering. The
models give a good description of the Tevatron data on production of $W$,
bottom and  jets in diffractive events, as well as jets with two rapidity 
gaps, alternatively leading particles. We also give predictions for diffractive
$J/\psi$ production and discuss diffractive Higgs production at the Tevatron 
and LHC.}

\begin{document}


Diffractive hard scattering provides through its hard scale (high $E_T$
jets, $W$ and heavy quark production) the perturbative parton level
basis for investigating non-perturbative effects. 
Diffraction, characterised by rapidity gaps and being a soft effect 
important on a long space-time scale, cannot be completely described
by perturbative QCD. 
Traditionally, the observation of rapidity gaps in hard scattering has been 
explained in terms of the pomeron model~\cite{IS}, which works well in 
describing the HERA data, but fails to reproduce the Tevatron diffractive 
data when parametrisations of the pomeron structure functions and pomeron 
flux from HERA are used. Another characteristic signature of diffractive 
scattering is the presence of a leading particle carrying most of the beam 
particle momentum ($x_F\gtrsim 0.9$), which is kinematically related to a 
rapidity gap. Within the pomeron picture, the leading particle 
spectrum is given entirely by the pomeron flux, which has been shown to
be non-universal.

A different approach is represented by the soft colour interactions models, 
based on the variations in the topology of the confining colour force fields: 
the soft colour interaction (SCI) model~\cite{SCI} and the generalized 
area law (GAL) model~\cite{GAL}. The SCI model is based on the assumption that 
partons emerging from the hard interaction can exchange colour with the proton
colour field in which they propagate. In GAL, the interaction between 
overlapping strings is the basis for colour exchange. Both models lead to 
rearrangements of string topology, in the former with the probability $P$ to 
exchange a soft gluon between pairs of partons, in the latter with the 
probability for two strings to interact given by $P=P_0 [1-exp(-b~\Delta A)]$,
based on the Lund string model.

\TABLE[t]{
\caption{Ratios diffractive/inclusive for hard scattering processes 
in \ppbar \ collisions at the Tevatron, showing experimental results
from CDF and D\O~\cite{CDF-W,CDF-JPSI,CDF-DPE} compared to the SCI and 
GAL soft color exchange models.
{\it Legend:} $^\mathrm{a}$ No result available;  $^\mathrm{b}$ Ratio 
of two-gap  events to one-gap events.}
\label{tab-gapratios} 
\begin{tabular}{lcclccc}
\hline
\hline
Observable & $\sqrt{s}$ &            & \multicolumn{3}{c}{Ratio [\%]} \\
           & [GeV]      & Experiment & Observed & SCI  & GAL  \\ 
\hline
$W$  - gap       & 1800 & CDF & $1.15 \pm 0.55$ & 1.2 & 0.8 \\
$Z$  - gap       & 1800 & --- &  ---$^\mathrm{a}$  & 1.0 & 0.5 \\
$b\bar b$  - gap & 1800 & CDF & $0.62 \pm 0.25$ & 0.7 & 1.4 \\
$J/\psi$  - gap  & 1800 & CDF & $1.45 \pm 0.25$ & 1.4  & 1.7 \\
$jj$ - gap       & 1800 & CDF & $0.75 \pm 0.10$ & 0.7 & 0.6 \\ 
$jj$ - gap       & 1800 & D\O & $0.65 \pm 0.04$ & 0.7 & 0.6 \\ 
$jj$ - gap       & ~~630& D\O & $1.19 \pm 0.08$ & 0.9  & 1.2 \\ 
gap - $jj$ -gap~$^\mathrm{b}$& 1800 & CDF  & $ 0.26 \pm 0.06 $ & 0.2 &0.1\\
$\bar{p}$ - $jj$ -gap~$^\mathrm{b}$ & 1800 & CDF  & $0.80 \pm 0.26$ &0.5& 0.4 \\
\hline
\hline
\end{tabular}
}

%

Both models have been implemented in Monte Carlo event generators, 
\Lepto~for $ep$ and \Pythia~for $p\bar{p}$ collisions.
Using the standard Lund hadronisation the models lead to different hadronic 
final states, giving rise to events with or without gaps, leading protons or 
neutrons, etc. Thus they provide a unified description of all final states.
They give a good description of rapidity gap events observed at 
HERA \cite{SCI,sce-heramc}, and good results when applied to 
hard diffractive \ppbar~ scattering at the Tevatron~\cite{ours}.
An overall summary of the relative rates of various diffractive hard processes 
is given  in Table~\ref{tab-gapratios}, which shows that this approach can 
account  for several different gap phenomena. It is  noteworthy that the SCI 
model  also reproduces the observed rate of high-$p_\perp$ charmonium and 
bottomonium  at  the Tevatron \cite{SCI-onium}, as well as charmonium
production in fixed target hadronic interactions \cite{Cristiano}.

The detailed description of the two soft colour models was presented
at lenght in \cite{ours}. The only 
parameter of the models, represented by the colour exchange probability
$P$, has been kept as fitted from diffractive DIS, although it was shown
that the diffractive ratios are not sensitive to the exact value.
Furthermore, for the case of \ppbar~ scattering, the models include 
treatment of the remnants of the colliding hadrons and description
of the soft underlying event, as developed for standard Monte Carlo
programs, like \Pythia.

An illustration of the effects of soft colour interactions, as well as  
a motivation for such modelling of the poorly known nonperturbative processes, 
is presented in Fig.~\ref{plotmaxgapsize}. The very strong effect
of hadronisation causes the large rapidity gaps present at the parton level
(dashed curve), to be exponentially suppressed (dashed-dotted curve) in the
hadronic final state after standard hadronisation. Applying the SCI model leads
to an increased probability for large rapidity gaps (full curve), which is
still far below the parton level result, but represents exactly what is needed
to describe the data. The appearance of a `diffractive plateau', taken
as a characteristic for diffraction, is visible
only when the kinematical constraint of the $W$ mass is relaxed (dotted
curve, $m_W=8$~GeV).


An experimental signature for diffraction is  given by
rapidity gaps. Events exhibiting one rapidity gap and having a hard scale ($W$,
\bbbar, high-$E_T$ dijets) have been observed at the Tevatron \cite{CDF-W}.
We find that soft colour models give a good description of the relative rates of
such events \cite{ours}, i.e. the ratio between the cross-section for a diffractive
process and the total cross-section for the same hard process (see Table
\ref{tab-gapratios}).

\EPSFIGURE[hr]{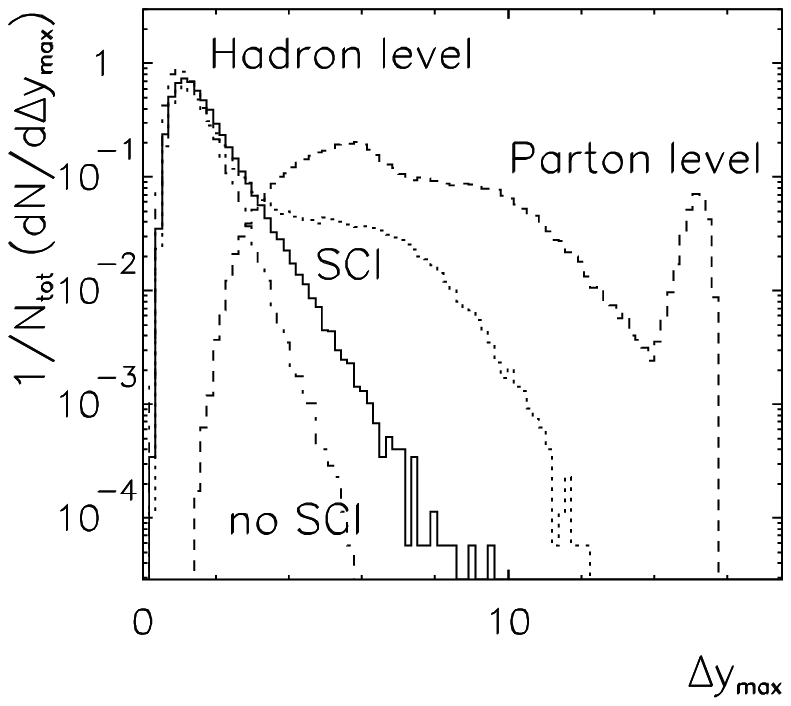,width= 0.45\textwidth,clip=}{Distribution of the 
size $\Delta y_{max}$ of the largest rapidity gap in $W$ production in \ppbar 
\ events at $\sqrt{s}=1.8$ TeV in \Pythia .\label{plotmaxgapsize}}

A different signature of diffraction is a leading proton. The 
observation of leading antiprotons at the Tevatron \cite{CDF-AP}, in events
with high-$E_T$
jets, offered a new testing ground for the models. The agreement with data
is good \cite{ours}, when comparing the total ratio of diffractive
to nondiffractive events, kinematical distributions of the dijets, and
the dependence of the ratio with the momentum  fraction $x$ of the interacting 
parton in $\bar{p}$. We find, however, an increased sensitivity
of the latter with the details in the modelling of the remnant.

It is the events with {\em two} leading protons with associated gaps that
will provide the ultimate test for the models. Such events, bearing the
traditional name double Pomeron exchange (DPE) because of their description 
in the  Regge framework, occur naturally in the soft colour interaction models. 
With one single mechanism for soft exchanges, the final colour string topology
may give rise to two rapidity gaps, or two leading protons. Such Double leading
Proton Events, with a dijet system in the central region, have been observed by
CDF~\cite{CDF-DPE}  and D{\O}.

\EPSFIGURE[hl]{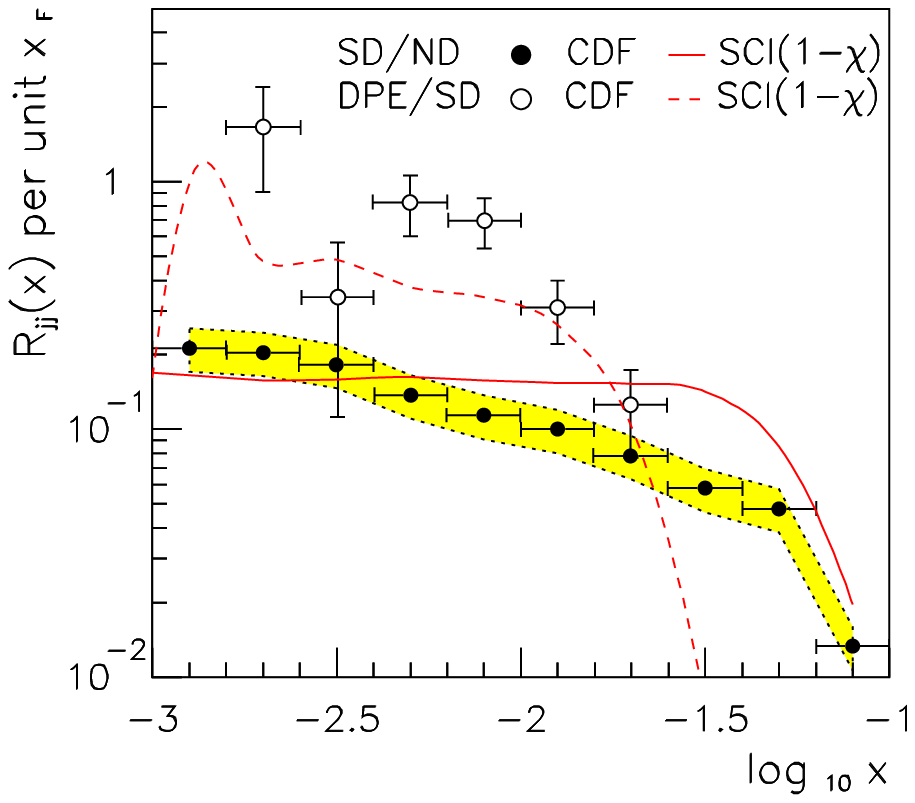,width= 0.45\textwidth,clip=}{The ratio of DPE 
to single diffraction (per unit $x_F^p$) and the ratio of single diffraction 
to nondiffraction (per unit $x_F^{\bar p}$), as a function of the momentum 
fraction $x$ of the  struck parton in $p$
and $\bar{p}$, respectively.\label{plot_r_dpe}}

Apart from reproducing quite well the jet properties in inclusive nondiffractive
(ND), single diffractive (SD) and DPE events, the models give also a reasonable
description of the relative ratios between these types of events, and
their cross-sections. For example, Fig. \ref{plot_r_dpe} gives a comparison
of the CDF data  with  the SCI model using a certain parametrisation for the
remnant treatment. Having two leading particles implies an increased 
sensitivity to the remnant treatment, providing possibilities to test and 
improve the details of the model.

The soft color interactions give rise to an even more striking effect of
two different phenomena in the same event,
namely both a rapidity gap and turning a colour octet \ccbar \ pair into a
singlet producing a $J/\psi$. It is a highly nontrivial result to explain
both these effects with one model for non-pQCD dynamics. The predictions
of our models are in agreement with the very recent observation by 
CDF \cite{CDF-JPSI} of such diffractive $J/\psi$
events (cf. Table \ref{tab-gapratios}).


Assuming a standard model Higgs exists,
Higgs production in diffractive hard scattering has been argued to be useful
for its discovery because of the Higgs signal standing out more
clearly in the cleaner diffractive events. This holds especially for Higgs 
production in DPE events, where the underlying event has exceptionally low 
activity. 

\TABLE[br]{
\caption{Cross sections and ratios for diffractive Higgs
production at the Tevatron and LHC for the soft 
color interaction models (using leading proton definition).}
\label{tab-higgs} 
\begin{tabular}{lcc}
\hline
\hline
$m_H=115$  &  $\sqrt{s}=1.96$~GeV  &  $\sqrt{s}=14$~GeV \\
   & SCI/GAL & GAL/SCI \\
\hline
$\sigma_{tot} [fb]$ & 600 & 27000 \\
$R^{SD}_H [\%]$    & 0.2  & 0.6 - 0.7 \\
$\sigma^{SD}_H [fb]$  & 1.2 & 162 - 189 \\
$R^{DPE}_H [\%]$   & 0.01 - 0.02 & 0.1 \\
$\sigma^{DPE}_H [fb]$  & $1.2 - 2.4 \cdot 10^{-4}$ & 0.162 - 0.189 \\
\hline
\hline
\end{tabular}
}

The existing predictions of the cross sections for these process vary by
several orders of magnitude, so the central question in judging whether this is
a useful Higgs channel, is whether the cross section is large enough.
In contrast to all other available theoretical calculations of
diffractive Higgs, our models have been tested against the available
diffractive data from the Tevatron. This puts us in a better
position to answer the question whether the diffractive Higgs channel is a
feasible one at the Tevatron and at LHC \cite{higgstobe}.

Table  \ref{tab-higgs} shows our predictions for diffractive
Higgs production, with a chosen mass of $m_H=115$ GeV, in events
associated with a leading proton (SD) or with two leading protons (DPE).
With the luminosity to be achieved at the Tevatron in Run~II, we conclude
that only a few single diffractive Higgs events can be observed, and no DPE events.
On the other hand, at the high CMS energy and luminosity available at the LHC,
diffractive Higgs physics can be thoroughly studied, and DPE Higgs events
will occur.


\EPSFIGURE[hl]{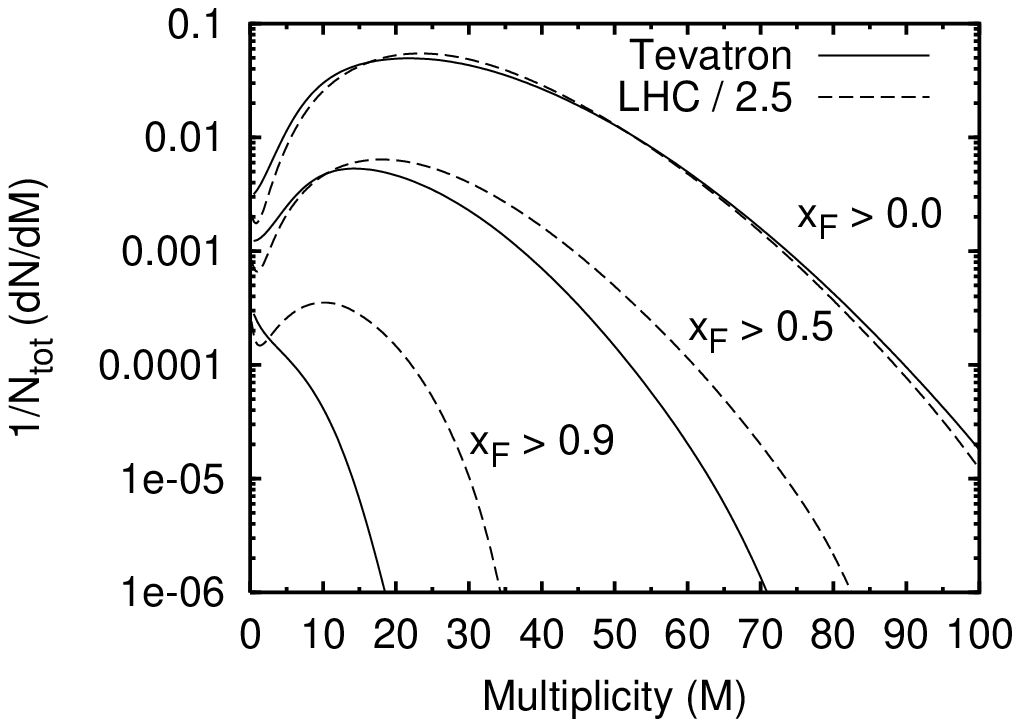,width= 0.45\textwidth,clip=}{Multiplicity 
in the hemisphere of a leading proton with minimum $x_F$, for Higgs events
from the SCI model in \Pythia.\label{multip}}

The quality of a diffractive event changes, however, at LHC energies.
Besides the production of a hard subsystem and one or two leading protons,
the energy is still enough for populating the expected rapidity 
gap regions with particles. As seen in Fig. \ref{multip}, the multiplicity
of particles in the rapidity region adjacent to the leading proton is considerably
higher at the LHC, compared to the Tevatron. Thus, a `clean' diffractive
signal for Higgs may not be achieved without paying the price of a lower
cross-section. Requiring one or two gaps induces a drop in the cross-section
by a factor 10, respectively 100, compared to Table \ref{tab-higgs}.

A new kind of rapidity gap events has been observed in \ppbar~collisions at the
Tevatron \cite{D0-jgj}, with a central rapidity gap spanned between two
high-$E_T$ jets, corresponding to a large momentum transfer across the gap. 
The soft colour  models cannot give a satisfactory description
of this phenomenon, which needs to be explained in terms of a hard colour
singlet exchange. Using a complete solution to the BFKL equation, which includes 
formally  non-leading corrections (consistency constraint, running $\alpha_S$),
the elastic parton-parton scattering amplitude
via colour singlet resummed gluon-ladders can be obtained \cite{jgj}. 

\EPSFIGURE[hr]{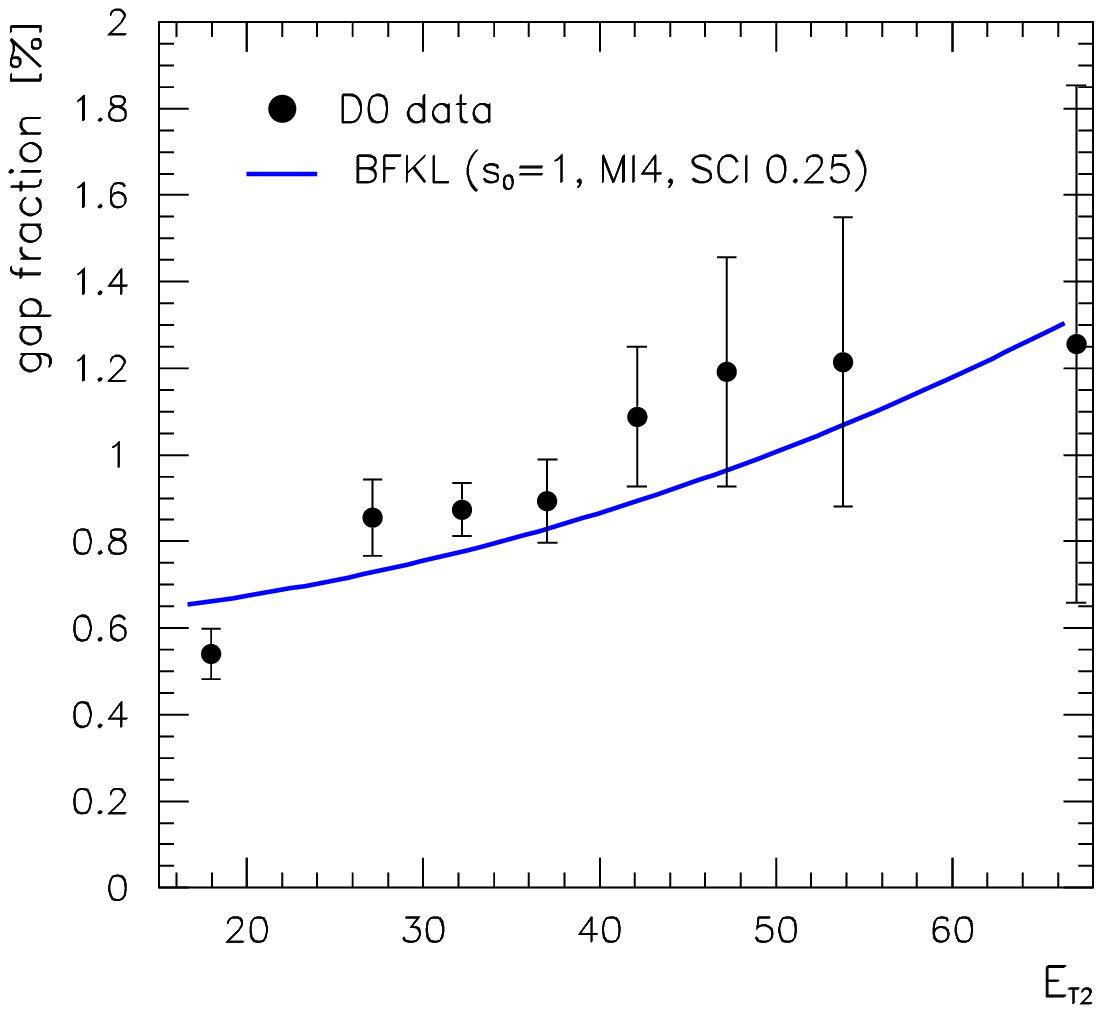,width= 0.45\textwidth,clip=}{Fraction of jet events having a rapidity
gap in $|\eta|<1$ between the jets, versus the second highest
jet-$E_T$.\label{jetgapjet}}

All soft effects, related
to producing the hadronic final state, can now be taken into account, by 
implementing the BFKL solution into \Pythia. As seen in Fig. \ref{jetgapjet},
the BFKL solution gives a good description of the D\O~data, with the Soft 
Colour Interaction model and Multiple Interactions
providing a gap survival
probability which varies event-by-event.


Our studies of the soft colour interaction models have demonstrated that
they are able to reproduce a wide range of hard diffractive phenomena, at  
HERA and the Tevatron, from low scales ($J/\psi$) to high scales ($W$). 
Thus, the soft colour mechanism plays an important role in understanding 
the hadronic final state.
Applying the models to diffractive Higgs production, we find cross-sections 
that are uninterestingly small at the Tevatron, but significant at the LHC.




\begin{thebibliography}{99}

  
\bibitem{IS} G.\ Ingelman and P.E.\ Schlein, \plb{152}{1985}{256}. 

\bibitem{SCI}
A.\ Edin, G.\ Ingelman, J.\ Rathsman, \plb{366}{1996}{371}; \zpc{75}{1997}{57}.

\bibitem{GAL}
J.\ Rathsman, \plb{452}{1999}{364}.


\bibitem{CDF-W}  CDF Collaboration, \prl{78}{1997}{2698};
\prl{84}{2000}{232}; \prl{79}{1997}{2636};
D0 Collaboration, \hepex{9912061}.

\bibitem{CDF-JPSI} CDF Collaboration, \hepex{0107071}.

\bibitem{CDF-DPE} CDF Collaboration, \prl{85}{2000}{4215}.

\bibitem{sce-heramc} A.\ Edin, G.\ Ingelman, J.\ Rathsman, \hepph{9912539}.

\bibitem{ours} R.~Enberg, G.~Ingelman, N.~Timneanu, \prd{64}{2001}{114015}.

\bibitem{SCI-onium} 
A.\ Edin, G.\ Ingelman, J.\ Rathsman, \prd{56}{1997}{7317}.

\bibitem{Cristiano} 
C.\ B.\ Mariotto, M.\ B.\ Gay Ducati, G.\ Ingelman, \hepph{0008200}. 

\bibitem{CDF-AP} CDF Collaboration, \prl{84}{2000}{5043}.

\bibitem{higgstobe} R.~Enberg, G.~Ingelman, N.~Timneanu, in preparation.


\bibitem{D0-jgj} D0 Collaboration, \plb{440}{1998}{189};
CDF Collaboration,\prl{80}{1998}{1156}.
\bibitem{jgj} R.~Enberg, G.~Ingelman, L.~Motyka, \hepph{0111090}.


\end{thebibliography}
\end{document}